\documentclass{PoS}

\title{Soft QCD Results from CMS}

\ShortTitle{CMS Soft QCD}

\author{\speaker{Yuan CHAO}\thanks{On behave of CMS Collaboration.}\\
        National Taiwan University\\
        E-mail: \email{yuanchao@cern.ch}}

\abstract{Studies of hadron production in proton-proton
collisions, including charged particle transverse momentum, pseudo-rapidity
and event-by-event multiplicity distributions at $\sqrt{s} =$ 0.9, 2.36 and
7 TeV are shown. Measured spectra of identified strange hadrons,
reconstructed based on their decay topology, are also presented.
Comparisons to several QCD Monte Carlo models and tunes are discussed.
Results on two-particle angular correlations over a broad range of
pseudo-rapidity and azimuthal angle in p-p collisions are presented at both
$\sqrt{s} = $ 0.9 and 7 TeV. In high multiplicity events, a pronounced
structure emerges in the two-dimensional correlation function for particle
pairs with intermediate transverse momentum of 1-3 GeV/$c$. Furthermore,
Bose-Einstein correlations between identical particles are measured in samples
of proton-proton collisions at $\sqrt{s} = $0.9 and 7 TeV. Finally,
a measurement of the underlying activity in scattering processes with a $p_T$
scale in the several GeV region is also presented.}

\FullConference{The 2011 Europhysics Conference on High Energy Physics,
                EPS-HEP 2011,\\
		July 21-27, 2011\\
		Grenoble, Rh\^one-Alpes, France}

\begin{document}

\section{Introduction}

The frontier energy scale of the Large Hadron Collider (LHC) provides us a
new tool to probe the physics at the new energy regime. With the general
purpose Compact Muon Solenoid (CMS) detector~\cite{CMS_det}, one can study
the underlying physics as well as have better constraint on the
phenomenological models with various
quantum chromodynamics (QCD) basic properties measurements. These studies are
essential to the searches of Higgs, SUSY and new physics as they provide
solid foundations to the understanding of the backgrounds.

The properties of proton-proton collisions are typically described either
empirically or with phenomenological models. For the Monte Carlo (MC)
simulation, event generator models are based on the assumption of parton
hadronization produced via the fragmentation of color strings. These models
are based on a convolution of parton distribution functions, the
hard-scattering cross section from perturbative calculations, and
fragmentation functions. PYTHIA~\cite{PYTHIA} and other general MC
generators further introduce multiple parton interaction (MPI) to account
for these contributions. The parametrization models as well as different
showering schemes are the MC ``tunings'' to be verified with experimental
results at the highest energy to date.

\section{Low $p_T$ QCD Measurements}

The soft QCD studies at CMS make use of the measurements on charged particles.
The full solid state tracking system of CMS gives a great niche on these
studies for its precision. The CMS tracker consists of 1440 silicon pixel
and 15148 silicon strip detector modules, immersed in the uniform 3.8 T
magnetic field provided by a 6 m diameter super-conducting solenoid.
The $p_T$ resolution for 1 GeV/$c$ charged particles is between $0.7\%$ at
$|\eta| = 0$ and $2\%$ at $|\eta | = 2.5$.

One can classify the soft collisions into: elastic scattering, inelastic
single-diffractive (SD) dissociation, double-diffractive (DD) dissociation,
and inelastic non-diffractive (ND) scattering~\cite{soft_int}. The interaction
events recorded by CMS are highly dominated by non-single-diffractive (NSD)
ones for the CMS trigger picks up mainly ND and DD events and disfavors SD
events.

Measurements of primary charged hadron transverse momentum ($p_T$),
pseudo-rapidity ($\eta$), and the multiplicity distributions ($N_{\rm ch}$)
are presented for NSD events in proton-proton collisions at centre-of-mass
energies $\sqrt{s} =$ 0.9, 2.36, and 7 TeV~\cite{PT, ETA, NCH}.
The phase-space-invariant differential yield $E d^3 N_{\rm ch} / dp^3$ for
primary charged particles is compared with an empirical parametrization with
the variable $x_T \equiv 2 p_T / \sqrt{s}$. The CMS results are consistent over
the accessible range in $x_T$.
The ProQ20 tune of PYTHIA 6 shows most consistence at $\sqrt{s} =$ 0.9 TeV,
while the 7 TeV data are most consistent with PYTHIA 8 at the $10\%$ level of
the full $p_T$ range. The measurements of $d N_{\rm ch} /d\eta$ distributions
are obtained, as in Ref.~\cite{ETAold}, with three methods, based on counting
the following quantities: (i) reconstructed clusters in the barrel part of the
pixel detector; (ii) pixel track-lets composed of pairs of clusters in
different pixel barrel layers; and (iii) tracks reconstructed in the full
tracker volume. The third method also allows a measurement of the
$d N_{\rm ch} /d p_T$ distribution.
The results for the three individual layers within the cluster-counting
method are found to be consistent within $1.2\%$ and combined. Also, the
three different measurement methods with agreement to the average within
$1\%$ to $4\%$ depending on $\eta$ are combined and compared to the
measurements of other experiments. The combined result for the central
pseudo-rapidity density is higher than most predictions and provides new
information to constrain ongoing improvements of theoretical models and
event generators. Independent emission of single particles yields a
Poissonian $P_n$, whose magnitude increases with $\sqrt{s}$~\cite{MP1}.
The mean of the multiplicity distribution, $\langle n \rangle$, is equal
to the integral of the inclusive single-particle density. The averaged
transverse momentum of the charged particles, $\langle p_T \rangle$, shows
a positive correlation with the event multiplicity. Traditionally,
one studies the Koba-Nielsen-Olesen (KNO) scaling~\cite{KNO} by testing
the KNO function $\Psi(z) = \langle n \rangle P_n$, where
$z = n/ \langle n \rangle$, on its $s$ dependence and the normalized
moments $C_q = \langle n^q \rangle / \langle n \rangle^ q$.
From the 7 TeV result, the change of slope in $P_n$ combined with the
strong linear increase of $C_q$ indicates a clear violation of KNO scaling
with respect to lower energies. The productions of $K^0_S$, $\Lambda$ and
$\Xi^-$ particles at $\sqrt{s} =$ 0.9 and 7 TeV~\cite{strange} are also
studied. The production ratios $N(\Lambda)/N(K^0_S)$ and $N(\Xi^-)/N(\Lambda)$
versus rapidity and transverse momentum show no change with the increase
of $\sqrt{s}$.
%centre-of-mass energy.

Bose-Einstein correlations have been measured for the first time at the LHC
with CMS in p-p collisions at $\sqrt{s} =$ 0.9, 2.36 and 7 TeV~\cite{BEC}.
The effective size of the emission region is observed to grow with
charged-particle multiplicity and to decrease with $k_T$ in events with
large multiplicity. On the other hand, long-range azimuthal correlations for
$2.0 < |\delta \eta| < 4.8$ have been studied for 7 TeV data~\cite{corr},
leading to the first observation of a long-range ridge-like structure at the
near-side ($\delta \phi \approx 0$) in p-p collisions.

\section{Underlying Event Study}
The ``Underlying Event'' (UE) is everything in a single proton-proton
interaction except for the hard scattering component. The UE contributes
through MPI as well as beam-beam remnants, concentrated along the beam
direction. It is important to have good understanding on the the UE
properties for either precision measurements of standard model processes
or searches for new physics at the frontier energy scale.

The first measurement of UE activity at the LHC has been published with CMS
data at $\sqrt{s} = $ 0.9 TeV~\cite{UE1}. MPI activity is expected to
increase with the centre-of-mass energy~\cite{MPI1, MPI2}. At a given
$\sqrt{s}$, the UE activity is also expected to increase with the interaction
hard scale. It will reach a plateau for high scales, which corresponds to an
MPI ``saturation'' effect. The recent published result~\cite{UE2} from CMS
measures the UE properties at $\sqrt{s} = $ 7 TeV with the same analysis
procedure and updates the 0.9 TeV results with a 30 times larger sample.
In this published result, all measurements are fully corrected for detector
effects. The UE studies are done with the referencing leading ''track-jet'',
reconstructed with charged particles. The activities in transverse region
in azimuthal direction, $60^\circ < \delta \phi < 120^\circ$, are most
sensitive to UE. The $p_T$ of the leading track-jet is taken as defining
the hard scale in the event. The large increase of activity in the transverse
region is observed in $N_{\rm ch}$, $\Sigma p_T$ and the the charged particle
$p_T$ spectrum up to $N_{\rm ch} = $ 30, $\Sigma p_T = $ 35 GeV/c, and
$p_T = $ 14 GeV/c. The results are compared with various tunes of PYTHIA 6
as well as PYTHIA 8 4C. A good description of distributions at $\sqrt{s} =$
7 TeV and $\sqrt{s}$ dependence from 0.9 to 7 TeV is from tune Z1. Tune Z2 and
PYTHIA 8 4C are also in reasonable agreement with data.

\section{Conclusion}
Comprehensive studies on various soft QCD topics have been done with CMS data
collected at $\sqrt{s} = $ 0.9, 2.36 and 7 TeV. UE analyzed at $\sqrt{s} = $
0.9 and 7 TeV, unfolded results compared with various MC tunes. These studies
can help on the discrimination of different theoretical modeling as well as
Monte Carlo tunings at the LHC energy-scale era.

\end{document}